\documentclass[conference]{IEEEtran}
\IEEEoverridecommandlockouts

\usepackage{csquotes} 
\usepackage{array}
\usepackage[backend=biber, natbib=true, minnames=1, maxbibnames=10, minbibnames=6, style=numeric, sorting=none, firstinits=true, urldate=long, backref=false, bibstyle=ieee]{biblatex}

\AtEveryBibitem{
}

\usepackage{flushend}

\usepackage[toc,hyperfirst=false,acronym,nonumberlist]{glossaries}
\makeglossaries

\usepackage{amsmath,amssymb,amsfonts}
\usepackage{algorithmic}
\usepackage{graphicx}
\usepackage{textcomp}
\usepackage{xcolor}
\usepackage{booktabs}
\usepackage{color, colortbl}
\definecolor{lightGray}{rgb}{0.8,0.8,0.8}

\usepackage{soul}

\usepackage{url}

\def\BibTeX{{\rm B\kern-.05em{\sc i\kern-.025em b}\kern-.08em
    T\kern-.1667em\lower.7ex\hbox{E}\kern-.125emX}}

\title{Cybersecurity Pathways Towards CE-Certified Autonomous Forestry Machines}

\author{Mazen Mohamad, Ramana Reddy Avula, Peter Folkesson, Pierre Kleberger,\\ Aria Mirzai, Martin Skoglund, Marvin Damschen\\
Dependable Transport Systems, RISE Research Institutes of Sweden, Borås, Sweden\\
\texttt{\small\{mazen.mohamad, ramana.reddy.avula, peter.folkesson, pierre.kleberger,}\\
\texttt{\small aria.mirzai, martin.skoglund, marvin.damschen\}@ri.se}
}

\begin{document}

\newacronym[longplural=Certificate Authorities]{CA}{CA}{Certificate Authority}

\newacronym{AI}{AI}{artificial intelligence}
\newacronym{EU}{EU}{European Union}
\newacronym{IoT}{IoT}{Internet of Thing}
\newacronym[longplural=Safety of the Intended Functionalities]{SOTIF}{SOTIF}{Safety of the Intended Functionality}
\newacronym{AHS}{AHS}{Autonomous Haulage System}
\newacronym{GNSS}{GNSS}{Global Navigation Satellite System}
\newacronym{GPS}{GPS}{Global Positioning System}
\newacronym[shortplural=SoS]{SoS}{SoS}{System of Systems}
\newacronym{SAC}{SAC}{Security Assurance Case}

\maketitle

\begin{abstract}

The increased importance of cybersecurity in autonomous machinery is becoming evident in the forestry domain. Forestry worksites are becoming more complex with the involvement of multiple systems and system of systems. Hence, there is a need to investigate how to address cybersecurity challenges for autonomous systems of systems in the forestry domain.

Using a literature review and adapting standards from similar domains, as well as collaborative sessions with domain experts, we identify challenges towards CE-certified autonomous forestry machines focusing on cybersecurity and safety. 
Furthermore, we discuss the relationship between safety and cybersecurity risk assessment and their relation to AI, highlighting the need for a holistic methodology for their assurance.

\end{abstract}


\section{Introduction}
\label{sec:intro}

The evolution of technology has significantly propelled autonomous mobile machines towards product readiness, even in safety-critical domains like forestry.
These machines, equipped with an increasingly sophisticated array of sensors, communication technologies, and \gls{AI}, promise to revolutionize tasks within this domain like site preparation and planting \cite{hansson2024autoplant}, as well as collection and transportation of logs \cite{la2023exploring}.
While offering increased productivity and reduced environmental impact, transitioning these technologies from laboratory settings to real-world applications introduces substantial challenges concerning safety and cybersecurity.

In the \gls{EU}, the CE marking represents a manufacturer's declaration that their product complies with the \gls{EU}'s health and safety requirements.
This conformity is crucial for introducing autonomous forestry machinery to the market, as it must be demonstrably safe for interaction with the general public.
Regulation (EU) 2023/1230 \cite{eu-2023-1230} on machinery, effective from early 2027, marks a significant update to the preceding Directive 2006/42/EC \cite{2006-42-ec}.
This regulation encompasses new technologies, including autonomous mobile machinery, \glspl{IoT}, and \gls{AI}, with a particular emphasis on cybersecurity requirements. Many new and forthcoming regulations may also need to be considered, e.g., Cyber Resilence Act \cite{com-2022-454}, Data Act \cite{eu-2023-2854} and AI act \cite{com-2021-206}. Hence, the pathway to compliance is complex. Regulations establish the legal framework for safety and cybersecurity, but do not detail the methods for achieving conformity.
International standards set by organizations such as ISO and IEC provide technical specifications, safety criteria, and performance metrics that help companies comply with regulations.
Harmonization of these standards with regulations simplifies the assessment of conformity, allowing products to meet or exceed regulatory requirements through adherence to recognized standards.
Unfortunately, as of this writing, no standards have been harmonized with Regulation (EU) 2023/1230, and there is a conspicuous absence of specific standards for the forestry domain addressing the primary challenges in autonomous machine conformity assessment: reliance on complex sensors as well as \gls{AI} for safety-critical functions as well as for maintaining cybersecurity.
Given its complexity, it is outside the scope of this paper to give a complete picture of the regulatory and certification challenges.
Instead, we introduce our work towards CE-certified autonomous forestry machines within the EU project AGRARSENSE\footnote{\url{https://www.agrarsense.eu}}.

Our contributions are as follows:
\begin{itemize}
    \item overview of challenges towards CE-certified autonomous forestry machinery addressed within the AGRARSENSE EU project,
    \item short survey of cybersecurity within forestry, including identification of the specific characteristics of this domain, and
    \item overview of how safety and cybersecurity risk assessments interact and how a combined methodology would be characterized.
\end{itemize}
Finally, we discuss challenges in treating assessments of safety, cybersecurity, and \gls{AI} separately, and sketch potential ways forward using assurance.

The rest of the paper is structured as follows. In Section~\ref{sec:bg}, we provide a background on the certification of machinery. In Section~\ref{sec:challenges} we present challenges towards the certification of autonomous forestry machines and in Section~\ref{sec:cybersecInForestry} we provide a survey on cybersecurity in the forestry domain. In Section~\ref{sec:discussion} we discuss assurance and compliance in forestry and Section~\ref{sec:conc} presents the concluding remarks and future work.
\begin{figure*}
    \centering
    \includegraphics[width=0.7\linewidth]{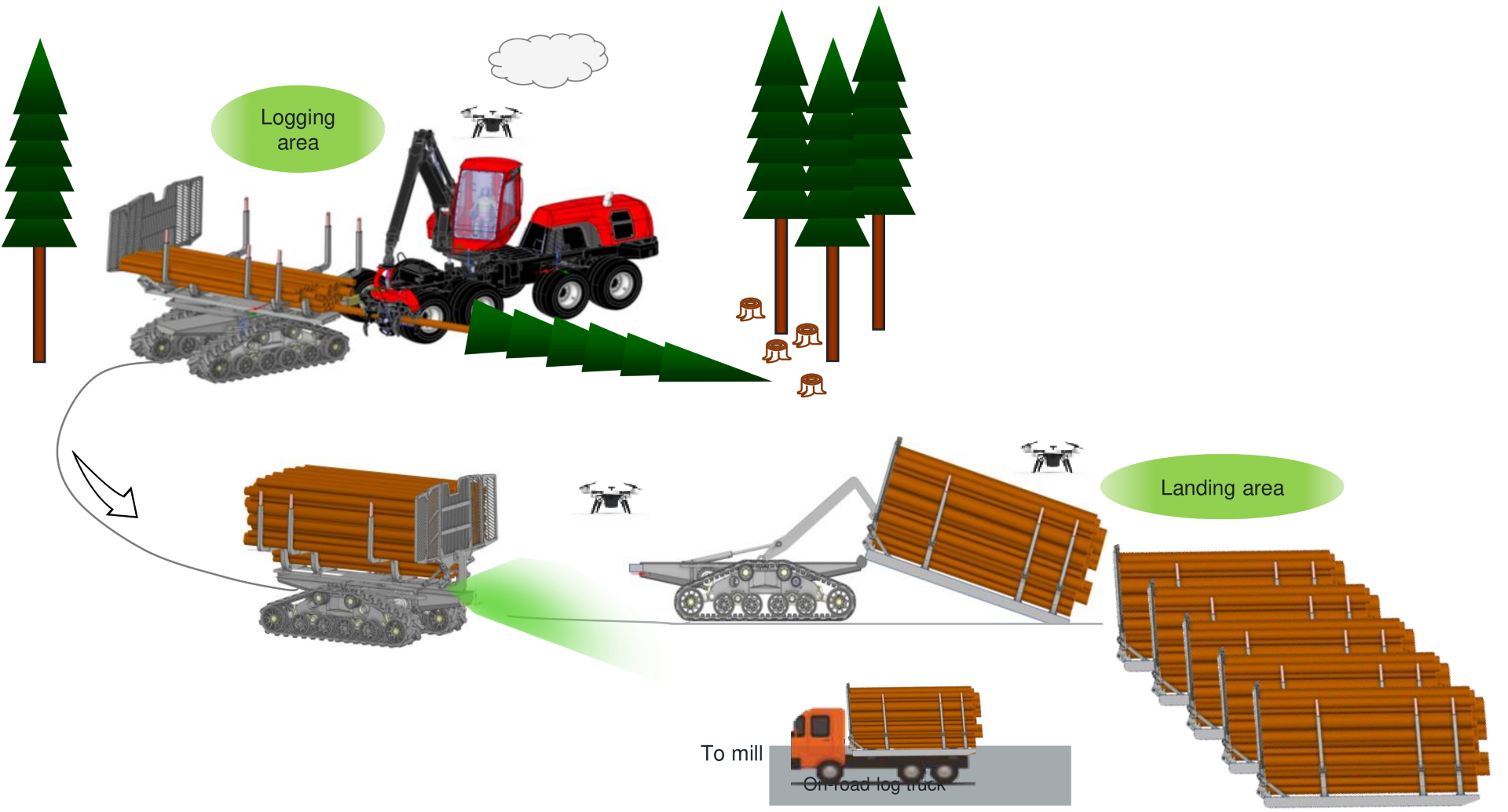}
    \caption{An illustration of the forestry worksite including autonomous forwarders, drones, and human-operated harvesting machines. \\Image courtesy of Komatsu Forest AB.}
    \label{fig:worksite}
\end{figure*}

\section{Background}
\label{sec:bg}
Historically, certification of machinery was focused on safety and compliance with mechanical and electrical standards.
However, with the advent of autonomous technology, the certification process is expanding to include software integrity, data security, and the ability of systems to make decisions in real-time without human intervention.
This shift necessitates a multidisciplinary approach to certification, involving expertise in cybersecurity, \gls{AI}, robotics, and even ethics \cite{Kusnirakova2023Rethinking}.
Today, the safety requirements for autonomous machines vary across different industrial sectors and countries, which reveals a gap between current standards and the state of technology \cite{Tiusanen2020An}.
One of the main challenges in designing a certification process for autonomous machines is not only the complexity of the systems itself but also the uncertainty about the changing environments in which the systems are deployed.
To address this, new certification frameworks have been proposed \cite{Fisher2018Verifiable}.

Our work contributes to bridging the gap within forestry machinery between current standards, emerging regulations and the state of technology.
While we address the general challenges of certifying autonomous forestry machines, our particular focus is on the intersection of cybersecurity and safety.
Our work is performed within the EU project AGRARSENSE.
Launched in January 2023, it aims to boost European agriculture and forestry productivity through innovative technologies.
It is coordinated by Komatsu Forest AB, involves 52 partners across 15 EU countries and has a total budget of approx. EUR 51 million over three years to address food security and climate challenges.

\section{Challenges Towards the Certification of Autonomous Forestry Machines}
\label{sec:challenges}
Within AGRARSENSE, we target the safety and cybersecurity of autonomous forestry machines.
More specifically, our research targets automation of transporting logs from a harvesting site to a landing area within the forest using an \textit{autonomous forwarder}, i.e., the forestry vehicle carrying the logs.
The aim is to increase productivity while reducing environmental impact.
It is assumed that harvesting itself is manually-operated, thus, the forestry worksite becomes partially autonomous.
Additionally, drones will be employed to observe the operations.
A key question we are investigating is how drones can complement safety-critical functions implemented on the autonomous forwarder, such as detecting people close to the machine.
An illustration of the envisioned worksite is depicted in Figure~\ref{fig:worksite}.

The critical need for safety and security in autonomous forestry machinery is emphasized by potential hazards including system failures, where machines might not detect obstacles, leading to collisions or catastrophic accidents. Furthermore, security breaches such as hacking could result in unauthorized machine operations, causing malfunction or unpredictable and dangerous behavior, thereby posing significant risks to operations and safety.
The aim of the project is to address safety and cybersecurity challenges holistically. In this section, we introduce the main challenges we are targeting.
Afterwards, we focus on our findings on cybersecurity within the forestry domain.


\subsection{Functional Safety of a Partially-Autonomous Worksite}
Functional safety is one of the key aspects to consider when demonstrating compliance with the requirements for CE marking in the forestry domain. The use of autonomous and manual machines working together imposes new risks that need to be assessed and, eventually, mitigated to ensure adequate levels of functional safety. In AGRARSENSE, some suggested mitigation strategies involve the use of collaborative safety functions such as a drone-based people detection function providing increased functional safety for the autonomous forwarders. Thus, novel risk assessment methods which consider both the interconnectedness and autonomy of collaborative systems are being investigated.
These risk assessment methods should integrate risk assessment strategies from standards belonging to different related domains, including machinery safety (ISO~13849~\cite{iso13849} and ISO~12100~\cite{12100}), automotive safety (ISO~21448~\cite{sotif}), and cybersecurity (IEC~62443~\cite{62443} and ISO/SAE~21434~\cite{iso21434}), adapting them to address specific challenges within the forestry domain.

\begin{figure}[h]
    \centering
    \includegraphics[width=0.9\linewidth]{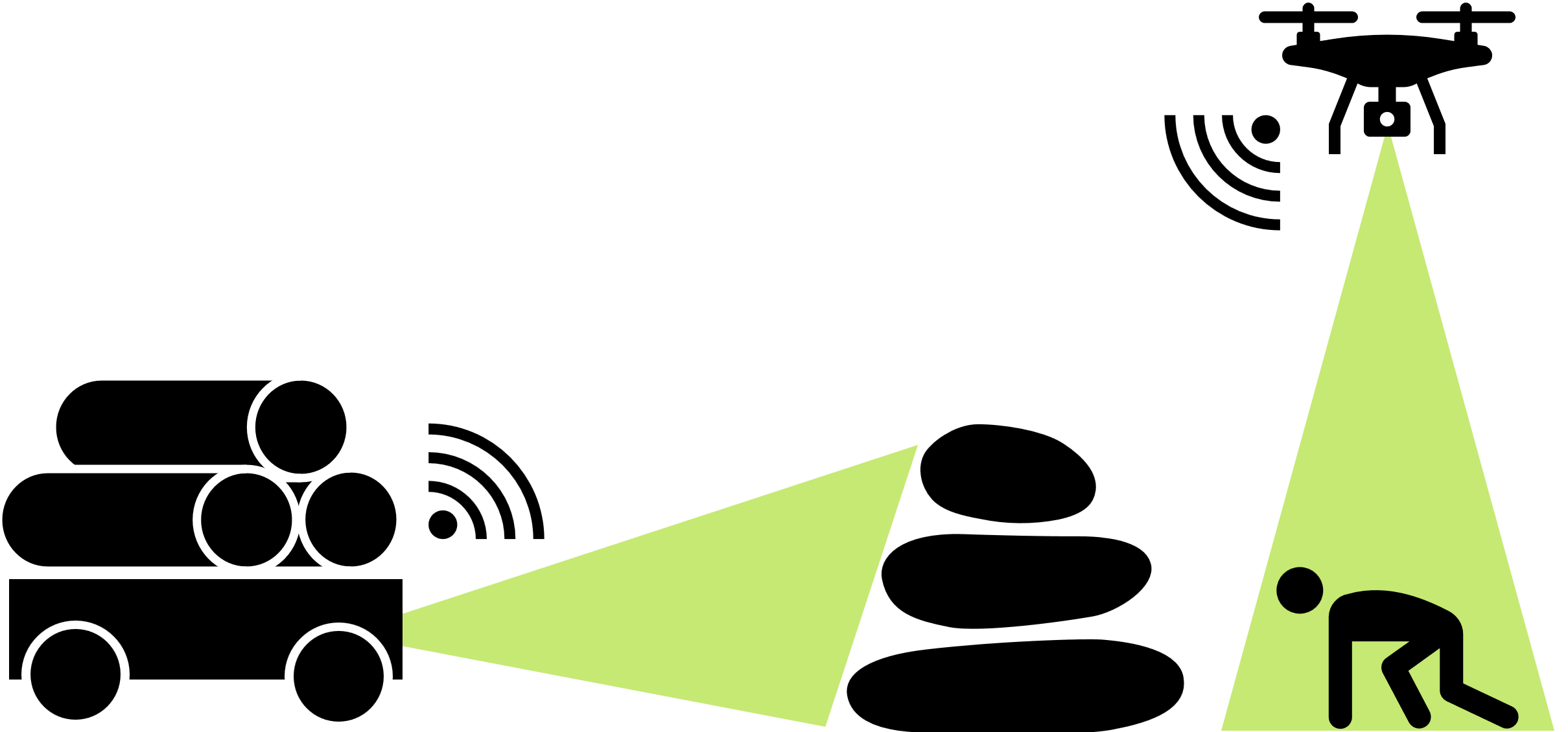}
    \caption{Use case description (minimalistic): The collaborative drone allows for an additional point of view to eliminate occlusions caused by terrain obstacles.}
    \label{fig:usecase}
\end{figure}

\subsection{Interplay between Safety and Cybersecurity}\label{sec:interplay}
The introduction of complex sensors and connectivity to enable automation in forestry elevates cybersecurity as a critical concern. With autonomous machines and interconnected systems becoming more widely deployed, there is an increased risk of security threats and system vulnerabilities.
Securing forestry automation systems is vital to ensure safety in addition to protecting sensitive information and maintaining operational continuity.
It is crucial to acknowledge that cybersecurity threats, e.g., attacks on communication, can potentially lead to unsafe behaviour in autonomous, connected vehicles \cite{malik2022comfase}.
Targeting the interplay between safety and cybersecurity involves ensuring that a system operates correctly and safely even in the face of cyber threats, thereby integrating safety measures with security strategies to protect against both accidental failures and intentional attacks.
Adopting an integrated approach to safety and cybersecurity is essential for addressing all relevant concerns \cite{Sync_MC}. This approach is an example of the contribution of EU-funded projects like AMASS, which have advanced the alignment of safety and security processes through synchronization points. 

\subsection{Safety of the Intended Functionality}
\label{subSeb:sotif}
Increased reliance on sensors leads to risks of non-hardware related functional inefficiencies like misinterpretation of sensor data, inadequate sensing due to environmental conditions or inadequate response to unforeseen situations.
In the automotive domain, safety measures address risks associated with a vehicle's intended functionality, e.g., automated emergency braking, using the \textit{\gls{SOTIF}} concept outlined in ISO~21448 \cite{sotif}.
Currently, no similar standard exists for machinery. Therefore, the AGRARSENSE project explores how to adapt \gls{SOTIF} principles to forest machinery and enhance safety beyond traditional functional safety standards like ISO~13849 \cite{iso13849}.
This work will consider the minimalistic use case depicted in Figure~\ref{fig:usecase}, where we will investigate whether risks due to insufficient situational awareness of the forwarder can be mitigated using an additional point of view. 

\subsection{Reliance on AI and Simulations}
Autonomous forestry machines are foreseen to rely on several \gls{AI} and machine learning components for vital tasks such as interpreting their surroundings using sensor data, performing object detection, and optimizing navigation paths through dense forest environments. The development of these \gls{AI} components requires vast amounts of data for training and validation. Unlike more active fields such as autonomous road vehicles, the forestry domain lacks comprehensive and diverse real-world data covering different operational scenarios and weather conditions. Creating such a dataset is challenging due to practical constraints such as access restrictions, environmental concerns, and low incentives for stakeholders. Advancements in graphics rendering and physics engines are increasingly making simulation data a crucial resource to supplement real-world data in the development of AI components \cite{nikolenko2021synthetic}.

Despite the apparent advantages of simulations in autonomous forestry machine development, one of the crucial challenges we are targeting is ensuring the validity and representativeness of the simulation data compared to the real world.
Addressing this challenge requires systematic validation of the components in the simulation toolchain in relation to the intended purpose. For example, assessing the validity of an \gls{AI} model for people detection trained using the simulation data would require validating the virtual sensor, simulated environmental factors such as lighting conditions or precipitation \cite{hasirlioglu2019general}, simulated human movement patterns, etc. Additionally, comprehensive validation procedures should include the evaluation of simulated terrain features, trees, weather dynamics, and the occlusions perceived by sensors due to obstacles.
While still under development, ISO/CD~PAS~8800~\cite{8800} as well as ISO/IEC~TR~5469~\cite{5469} can provide guidance in systematically developing, testing, and validating AI components of autonomous forestry machines, ensuring safety, reliability, and ethical considerations.

\section{Cybersecurity in Forestry}
\label{sec:cybersecInForestry}
\begin{figure}
    \centering
    \includegraphics[width=\linewidth]{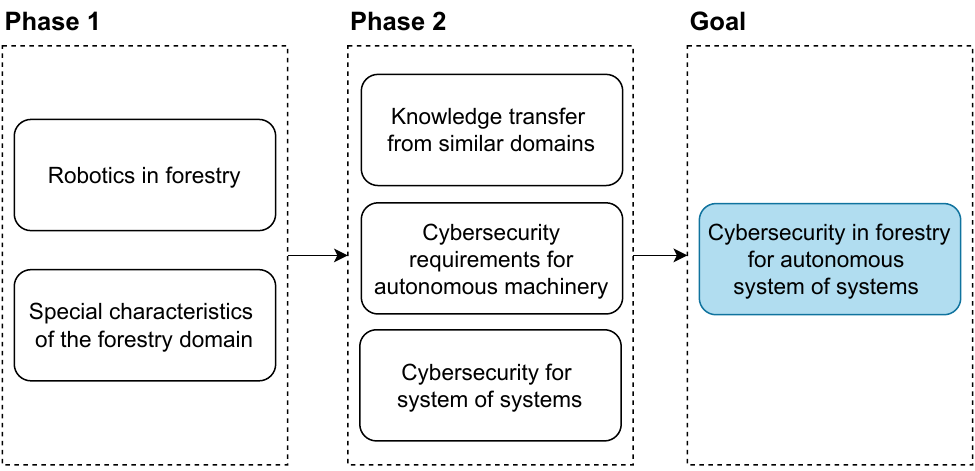}
    \caption{Approach for understanding cybersecurity for an autonomous system of systems in the forestry domain.}
    \label{fig:RW_approach}
\end{figure}

The previous section introduced the main challenges within safety and cybersecurity that are targeted within the AGRARSENSE project.
While the general need for cybersecurity and its relation to safety concerns was motivated in Section~\ref{sec:interplay}, the specific aspects of cybersecurity within the forestry domain still need to be identified.
In order to gain a solid understanding of the background and related work for cybersecurity in our use case, we used the approach depicted in Figure~\ref{fig:RW_approach}.

In the first phase, since our use case is in forestry and uses autonomous machines and robots, we started reviewing articles related to \emph{robotics in forestry}. The forestry domain has specific characteristics which we need to take into consideration when performing cybersecurity risk assessment. Hence, we looked into \emph{specific characteristics of the forestry domain}. Combining what we learned in the first two phases, we concluded that there is a lack of relevant literature on cybersecurity in forestry. Hence, we started the second phase by exploring possible \emph{knowledge transfer from similar domains}. We are also going to integrate different systems, e.g., an autonomous forwarder and a drone (see Figure~\ref{fig:usecase}), to achieve the goal of the use case, and for that, we investigated \emph{cybersecurity for system of systems}. Furthermore, since we are aiming to use autonomous machinery we had to gain an understanding of the \emph{cybersecurity requirements for autonomous machinery}. 

\subsection{Robotics in Forestry}
The use of robotics and autonomous machines in forestry is becoming more common. \citet{roboticsInForestry} review research articles and commercial products of robotic applications for different purposes, e.g., monitoring, wildfire firefighting, and harvesting.
The paper includes a review of multiple research and commercial projects that are relevant to the use case of this study, which is considering the use of autonomous machinery, e.g., drones and forwarders, in forests to perform tasks such as monitoring and inventory operations, as well as the specifications of these machines, e.g., sensors. However, the paper does not touch upon cybersecurity and safety concerns for the reviewed systems. Moreover, we reviewed the commercial products reported in the survey by visiting their websites, and could not find any relevant information regarding safety and cybersecurity aspects.

\citet{bergerman2016robotics} discuss robotics in agriculture and forestry. They state that the focus of academic and commercial research is on sensing, mobility, and manipulation technologies to enhance agriculture and forestry output and productivity. The paper includes numerous case studies in the area. However, again cybersecurity was not sufficiently represented in the report. 

Similarly, \citet{roldan2018robots} review the state of the art in robotics in agriculture and discuss automation in the field without any mention of the cybersecurity implications of such automation. 

\citet{abdelsalam} present a literature review to find the current practical autonomous navigation and material handling solutions that are suitable for the mill yard environment and what sorts of sensors are utilized in these systems.
While the report can serve as a good starting point for cybersecurity assessment of the relevant commonly used sensors, cybersecurity was not a characteristic included or analyzed in the literature review.

To summarize, cybersecurity is rarely studied in combination with forestry robots and autonomous machinery.

\subsection{Specific Aspects of Forestry}
To gain an understanding of the specific characteristics of the forestry domain in relation to cybersecurity, we performed a brainstorming session including 10 experts in cybersecurity \& safety (8), and forestry (2). The session started with presenting a pre-defined list of potential forestry-specific characteristics, which was then discussed, refined, and extended resulting in the characteristics identified and described in Table~\ref{tbl:forestrySpecifics}. These characteristics serve as the basis for determining the domain or domains from which a knowledge transfer would be considered.

\begin{table*}[!ht]
    \centering
    \caption{Specific characteristics of the forestry domain to be considered when performing cybersecurity analysis}
    \label{tbl:forestrySpecifics}
    \begin{tabular}{p{0.2\linewidth}p{0.75\linewidth}}
\toprule
    Characteristic & Description \\
\midrule
       Remote and Isolated Locations  & Many forestry operations occur in remote and isolated areas with limited connectivity. Ensuring secure communication and data protection in such environments can be challenging. \\
      \rowcolor{lightGray} Autonomous Machinery  & The use of autonomous machinery, such as drones and robots, is increasing in forestry. Securing these autonomous systems from cyber threats is crucial to prevent unauthorized access or interference.\\
       Natural Disasters  & Forestry operations can be susceptible to natural disasters like wildfires, floods, and storms. Cybersecurity measures should consider disaster recovery and business continuity planning to address cybersecurity issues that may arise during and after such events.\\
      \rowcolor{lightGray} Data Privacy and Compliance  & Forestry organizations may handle sensitive data related to land ownership, environmental impact assessments, and legal compliance. Ensuring data privacy and compliance with relevant regulations is critical to cybersecurity.\\
       Remote Monitoring and Control  & Remote monitoring and control systems are commonly used to manage equipment and collect data from remote forest locations. Securing these systems is essential to prevent unauthorized access and potential disruptions to operations\\
      \rowcolor{lightGray} Threat Profile  &  Creating threat profiles for companies in the forestry domain is important to grasp the potential threats, threat agents, and possible control measures.\\
       Confidentiality of Operations  & In some cases, e.g., military sites, operations in the forestry domain are confidential. Cybersecurity measures should ensure that the operations and corresponding communications are done in a confidential manner\\
        \rowcolor{lightGray} Heavy Machinery  &  The use of heavy machinery in forestry, e.g., harvesting machines, increases safety risks, and in turn increases cybersecurity concerns, particularly regarding threats that could compromise safety.\\
       \bottomrule
    \end{tabular}
\end{table*}

Forestry environments are inherently harsh and remote, presenting unique safety and cybersecurity challenges for autonomous systems. These settings lack the cooperative and connected functions that support similar technologies in more accessible areas. Unlike urban environments, where infrastructure and automated systems enable extensive communication and cooperation, forestry operations must rely on internal communications within a broader system of systems. This specific context necessitates tailored solutions designed to function effectively in the isolated and infrastructure-limited settings typical of forestry applications. For instance, limited connectivity alters the use of reactive and adaptive cybersecurity strategies in these settings.

\subsection{Cybersecurity in Similar Domains}
To the best of our knowledge, cybersecurity has very scarcely been studied in relevance to forestry. 
However, this is not the case in similar domains, such as the mining industry.

\citet{mining1} studied the relationship between the safety and cybersecurity of \glspl{AHS}, used in the mining industry for the transportation of ore autonomously and/or with remote control. 
The paper identifies and highlights challenges and open issues related to the cybersecurity and communication of \glspl{AHS} by conducting a literature survey.

\citet{mining1} found the main cybersecurity issues to be within the communication and its reliability. \glspl{AHS} depend on wireless communication between different components such as object avoidance/detection systems, \glspl{GNSS} (e.g., \gls{GPS}), and \gls{AI}.
The authors identify challenges related to these wireless communications, e.g., frequency interference when two devices send signals with similar frequencies to the same receiver; channel utilization to maximize the efficiency of the used channels; and signal jamming where attackers attempt to disrupt the communication by sending strong signals and noise.
Vulnerabilities also arise with wireless communication as discussed by the authors. These include: Wi-Fi De-Auth attacks to disconnect \gls{AHS} vehicles from the network, disrupting operations; \gls{GNSS} attacks to spoof or jam \gls{GNSS} signals, causing inaccurate navigation by \gls{AHS} vehicles; and camera attacks to steal video footage from \gls{AHS} vehicles or to control the vehicles' cameras remotely.

Automotive is another domain in which cybersecurity for autonomous vehicles has been studied.
\citet{mining2} discuss the security of autonomous vehicles and lists the various types of sensors that are used in such systems, e.g., \gls{GNSS}, LiDAR, Ultrasonic sensors, and cameras. The authors discuss possible attacks against these sensors and highlight possible defense strategies. Further, the study discusses if these defense strategies require modifications and extra hardware. For example, the defense strategies for \gls{GNSS} attacks can be checking the signals characters, e.g., strength, and applying cryptography in terms of encryption and modification.  

For camera-related attacks, the authors refer to the work of \citet{mining4} and the use of redundancy where multiple cameras cooperate, and special lenses such as photochromic lenses provide adequate protection from various angles against camera attacks. 

Other mitigation strategies against camera-related attacks are suggested by \citet{mining5}. These involve the usage of \gls{AI} to detect and mitigate remote attacks via a dedicated anti-hacking device.

\citet{mining3} approach autonomous vehicles from a cyber-physical system perspective and discuss the main cybersecurity challenges. The authors emphasize the importance of having a \gls{CA} in place to issue certificates to components involved in the communication with cyber-physical systems to avoid untrusted components from initiating attacks. 

To summarize, since our study focuses on autonomous machinery in forestry that relies heavily on wireless communication, we believe that the vulnerabilities and challenges identified and presented in the literature for the mining industry are of high relevance. Additionally, we found threats, vulnerabilities, challenges, and mitigation strategies that target autonomous vehicles to be relevant and can serve as a starting point in our approach to the cybersecurity of autonomous forestry machinery. 

\subsection{Cybersecurity Requirements in Related Standards}
In the broader context of machinery and automotive sectors, standards like ISO/SAE~21434 \cite{iso21434} and IEC~62443 \cite{62443} have been essential in defining cybersecurity requirements.
ISO/SAE~21434, focusing on-road vehicles, provides a structured approach to cybersecurity engineering throughout the lifecycle of the vehicle, emphasizing risk management, design, verification, and response strategies.
On the other hand, IEC~62443, dedicated to industrial communication networks and system security, offers a comprehensive framework for protecting industrial automation and control systems against cybersecurity threats.
Although originally intended for specific sectors, these standards present principles and methodologies adaptable to the cybersecurity needs of autonomous forestry machinery.

This fact is acknowledged by the technical report IEC~TS~63074 \cite{63074}, which details the use of the IEC~62443 standard in relation to safety-related control systems in the machinery domain.
It emphasizes the intersection of safety and cybersecurity, recognizing that security threats and vulnerabilities could potentially compromise the functional safety of safety-related control systems, thus impacting the safe operation of machinery.
It underscores the necessity for a comprehensive security risk assessment, aligned with IEC~62443, to identify and mitigate security threats that could affect these control systems.
Moreover, IEC~TS~63074  outlines specific security countermeasures and strategies, such as identification and authentication, access control, system integrity, and data confidentiality, among others, aimed at protecting machinery from unauthorized access and ensuring the integrity and availability of safety functions.

To conclude, cybersecurity requirements and countermeasures relevant to autonomous forestry machinery can be extracted from ISO/SAE~21434, IEC~62443, and guidance from IEC~TS~63074. 
However, this is non-trivial to do for developers and operators of autonomous forestry machinery wanting to enhance their systems' resilience against cyber threats while maintaining safety and operational integrity.
Thus, a forestry-specific standard providing a holistic approach to cybersecurity, referencing both general and machinery-specific standards, is desired to develop and deploy secure, safe, resilient, and reliable autonomous forestry machinery.

\subsection{Cybersecurity for System of Systems}
When conducting a cybersecurity assessment for a \gls{SoS}, it is essential to consider a wide range of factors and challenges. This is mainly because in \gls{SoS} we are connecting separate systems and components. Ensuring the security of individual elements is insufficient; rather, security must be assured for the integrated system as a whole.
\citet{sos} discuss the key cybersecurity problems of \gls{SoS} and these can be summarized as follows:
\begin{itemize}
    \item \textit{Operational Independence:} \gls{SoS} components operate separately, with varying policies, technologies and requirements, potentially causing conflicts. Vulnerabilities in some parts can jeopardize the overall security of the complete system.
    \item \textit{Management Independence:} Different organizations may manage different component systems which may introduce security concerns, as actions of one system might impact the security of others. 
    \item \textit{Evolutionary Development:} As \gls{SoS} evolves, new security issues may arise that were not initially anticipated. Security protocols and control measures need to evolve alongside the \gls{SoS} to address these emerging challenges.
    \item \textit{Emergent Behavior:} After deployment, \gls{SoS} behave and function in a non-localized manner. This can potentially lead to security issues. Determining responsibility for these distributed behaviors is intricate and shared among multiple entities, posing challenges for effective responses.
    \item \textit{Geographic Distribution:} The geographical spread of a \gls{SoS} complicates security endeavors as different national regulations can restrict coordination and timely responses.
\end{itemize}
Reflections on the listed problems highlight a fundamental challenge stemming from geographical distribution, operational independence, and managerial autonomy for the intended \gls{SoS}. These factors significantly contribute to the design's complexity and validation and verification processes. Ideally, these concerns should be proactively addressed and integrated into the initial design phase, extending throughout the entire product lifecycle. Moreover, this complexity is further compounded by the evolutionary development and emergent behaviors intrinsic to \gls{SoS}.
To summarize, \gls{SoS} cybersecurity issues involve various challenges and difficulties, including coordinating, detecting and responding, adapting security measures, and understanding the security posture across the \gls{SoS}.

\section{Discussion on Assurance and\\ Compliance in Forestry}
\label{sec:discussion}
We see moving from the challenges in certifying autonomous forestry machinery to assurance and compliance as a crucial step.
In the previous section, we outlined key challenges in safety, cybersecurity, and AI.
In the following, the challenges are tied to compliance strategies.
This approach aims to navigate the certification landscape efficiently, ensuring the autonomous forestry machinery meets the stringent safety and operational integrity criteria.

Cybersecurity assurance is important to gain confidence that a particular system has implemented the required cybersecurity measures to protect it from cyber threats.
One common approach for assurance is to create assurance cases that are structured bodies of arguments and evidence used to reason about a specific concern of the system. When the concern is cybersecurity, we create \glspl{SAC}. SAC can be represented in different ways, e.g., using the Goal Structure Notation (GSN) \cite{gsn}, or Claim Argument Evidence (CAE) \cite{CAE}. 

Although the main reason for creating \glspl{SAC} is to demonstrate compliance with regulations and standards, they can also be used in multiple usage scenarios, e.g., cybersecurity assessment, decision support, and in case of litigation \cite{mohamad2020security}. Modern assurance frameworks also have the potential to support innovation and continuous incremental assurance \cite{assurance2}.

Despite having many approaches for creating \glspl{SAC} reported in the literature, their industrial validation is limited \cite{SAC_SLR}.
The automotive domain is a step ahead when it comes to \glspl{SAC}, as it is explicitly required in relevant regulations and standards \cite{mohamad2020security}.

Forestry worksites present a complex environment consisting of multiple machines and systems essential for operations. 
Among these are autonomous machines that integrate \gls{AI} functions into their systems. 
Since these machines and systems collectively work towards a common goal, they can be considered a \gls{SoS}.
Hence, it is logical to coordinate the assurance of different system concerns, such as safety, cybersecurity, and \gls{AI} functions as suggested by Bloomfield et al. \cite{securityInformedSafety}.
However, the landscape is complicated by the existence of diverse regulations and standards governing the different properties. 
Hence, compliance requirements necessitate the separation of concerns, which calls for creating and adopting a modular approach for an assurance framework. 
For that, we want to do a knowledge transfer of an approach for creating \glspl{SAC} that has been evaluated in multiple domains \cite{mohamad2023cascade} and use it for forestry.
We intend to extend the approach to include arguments and evidence about safety and \gls{AI} regulations and standards requirements fulfillment. 

\section{Conclusion and Future Work}
\label{sec:conc}
In this paper, we laid out the primary cybersecurity challenges that need to be addressed in order for the AGRARSENSE EU project to pave the way for CE-certified autonomous forestry machines successfully. These include the need to adapt risk assessment strategies from relevant standards such as IEC~62443 \cite{62443} and ISO/SAE~21434 \cite{iso21434} to the forestry domain, allowing for the interplay between safety and security risk assessment and assessing the reliability of \gls{AI} and simulation data.
Our review of cybersecurity in the forestry domain revealed a scarcity in the reported literature and the need for further research considering the increasing complexity of forestry operations and worksites due to the introduction of autonomous machinery, system of systems, and \gls{AI}.

As future steps, we will be working on developing a forestry-adapted risk assessment methodology, using ISO/SAE~21434 (in particular the continuous risk assessment part), IEC~62443 (including the adaptation of the risk assessment method to various domains) and IEC~TS~63074 \cite{63074} as guidance. This methodology will take the interplay between safety and cybersecurity into consideration, meaning the prevention of emerging safety risks due to cybersecurity compromises. To our knowledge, no harmonised standard addressing safety and cybersecurity has yet been proposed. And just as in our use case, we predict the introduction of autonomous machinery to be correlated with an increased reliance on multiple interconnected systems, hence the methodology should also be applicable for \gls{SoS}. We believe that the more mature machinery usage in the mining sector can offer substantial guidance to accelerate this work. The developed method will be applied to assess the risks of the use case described in Figure \ref{fig:usecase}, Section \ref{subSeb:sotif}.

Additionally, we will develop a validation method for simulation environments to ensure that their obtained results possess an adequate representation of the real world. This will be a crucial requirement in line with the increasing integration of \gls{AI} components, as is the need for comprehensive and high-quality forestry datasets. 

Lastly, we will investigate introducing \glspl{SAC} to the forestry domain to gain confidence that a system has implemented the cybersecurity measures required to protect it from threats. This can be done through a knowledge transfer of approaches to build \glspl{SAC} from other domains.

\section*{Acknowledgment}
We thank Komatsu Forest AB for valuable discussions as the leader of the forestry use case.
AGRARSENSE is supported by the Chips JU and its members, including the top-up funding by Sweden, Czechia,
Finland, Ireland, Italy, Latvia, Netherlands, Norway, Poland and
Spain (Grant Agreement No. 101095835).
The views expressed in this document are the sole responsibility of the authors and do not necessarily reflect the views or position of the European Commission.
Neither the authors, the AGRARSENSE consortium, nor the Chips JU are responsible for the use which might be made of the information contained in this document.

\printbibliography[heading=bibintoc]

\end{document}